\documentclass[
	a4paper, 
	10pt, 
	twoside, 
]{LTJournalArticle}

\footertext{Published in \textit{Optics Express}. Submitted to Feb. 7, 2023. Accepted May 23, 2023.} 

\usepackage{breqn,floatrow,hyperref,appendix,float,enumitem,fancyhdr,siunitx,amsmath,amsfonts,setspace,gensymb,graphicx,caption,ulem}
\usepackage{dblfloatfix}

\title{Large-scale optical compression of free-space using an experimental three-lens spaceplate} 

\author{
Nicholas J. Sorensen, Michael T. Weil, and Jeff S. Lundeen
}

\date{\large\textit{Physics Dept. and Nexus for Quantum Technologies, University of Ottawa, 25 Templeton Street, Ottawa ON K1N 6N5, Canada}}

\renewcommand{\maketitlehookd}{%
	\begin{abstract}
		\noindent Recently introduced, spaceplates achieve the propagation of light for a distance greater than their thickness. In this way, they compress optical space, reducing the required distance between optical elements in an imaging system. Here we introduce a spaceplate based on conventional optics in a 4-$f$ arrangement, mimicking the transfer function of free-space in a thinner system - we term this device a three-lens spaceplate. It is broadband, polarization-independent, and can be used for meter-scale space compression. We experimentally measure compression ratios up to 15.6, replacing up to 4.4 meters of free-space, three orders of magnitude greater than current optical spaceplates. We demonstrate that three-lens spaceplates reduce the length of a full-color imaging system, albeit with reductions in resolution and contrast. We present theoretical limits on the numerical aperture and the compression ratio. Our design presents a simple, accessible, cost-effective method for optically compressing large amounts of space. 
	\end{abstract}
}

\begin{document}

\maketitle

\section{Introduction}
The miniaturization of optical systems, such as microscopes, telescopes, and cameras, has long been a topic of interest for physicists. By designing smaller optical systems, we can increase their versatility and reduce their cost. Most work has aimed at making the lenses thinner, e.g., metalenses \cite{Moon2022, Engelberg2020, Khorasaninejad2016, Liang2018, Chen2017, Overvig2019, Yu2014, Li2022}, diffractive lenses \cite{Sweeney1995, Faklis1995, Banerji2019, Meem2018, Kim2013, Overvig2022}, and Fresnel lenses \cite{Joo2022}. Instead, our aim is to miniaturize the space \textit{between} lenses, an often neglected element in optical systems. Since this is usually the largest contributor to the total length of an imaging system, its miniaturization promises the largest impact.

Recently, the "spaceplate" was introduced as an optical element that compresses the effect of propagation on light into a plate \cite{Reshef2021}. Such a device, by definition, must replace a slab of free-space with thickness $d_{\text{eff}}$ while only occupying a physical thickness $d$ less than $d_{\text{eff}}$. The compression ratio, $\mathcal{R}=d_{\text{eff}}/{d}$, is used to characterize spaceplate designs \cite{Reshef2021}. Like free-space propagation, a spaceplate changes the height of a ray but leaves its angle unchanged. Thus, unlike a lens, which does change ray angle, a spaceplate does not alter the magnification of an imaging system. From the view of physical optics, the spaceplate changes the phase of an incident plane wave without altering its angle. 

A variety of optical devices have been proposed to compress space. The work of Reshef et al. \cite{Reshef2021} first introduced and experimentally demonstrated the spaceplate. They experimentally tested two designs, one being a uniaxial crystal and the other a low-index medium. The designs were found to have compression ratios of $\mathcal{R}=1.12$ and $1.48$, respectively, which resulted in each spaceplate saving less than $\SI{4}{mm}$ of space. Remarkably, the uniaxial crystal had a large numerical aperture ($\text{NA}=\sin\theta$) and functioned at high incident angles $(\text{NA}=0.85)$ while remaining broadband, i.e., highly transmissive across the visible spectrum. In the same paper, the group also proposed and modeled a multilayer thin-film design. This had a compression ratio of $\mathcal{R}=5.4$, but was only capable of saving $\SI{50}{\micro m}$ of space and was narrowband. In another, almost simultaneous work, Guo et al. \cite{Guo2020} numerically demonstrated a much higher compression ratio of $\mathcal{R}=144$ using the Fano resonances of a photonic crystal slab at wavelength $\lambda$. Unlike the uniaxial crystal, however, this design was significantly limited in NA $(\text{NA}=0.01)$ and the actual space saved was limited to approximately $360\lambda$ \cite{Guo2020}. Subsequent designs, including those made of homogeneous materials \cite{Reshef2021}, improved multilayer stacks \cite{Page2022}, repeated Fabry-Perot cavities \cite{Chen2021, Mrnka2022}, and others \cite{Long2022, Chen2022, Ivanov2022}, made incremental improvements upon compression ratio and NA. None, however, can be scaled up to save space on the order of meters. Additionally, most of the previously demonstrated designs are limited in either polarization or bandwidth \cite{Reshef2021, Guo2020, Page2022, Chen2021}. In this paper, we present a broadband, polarization-independent three-lens spaceplate capable of meter-scale compression. 
\begin{figure*}[t]
     \centering
     \includegraphics[width=0.85\textwidth]{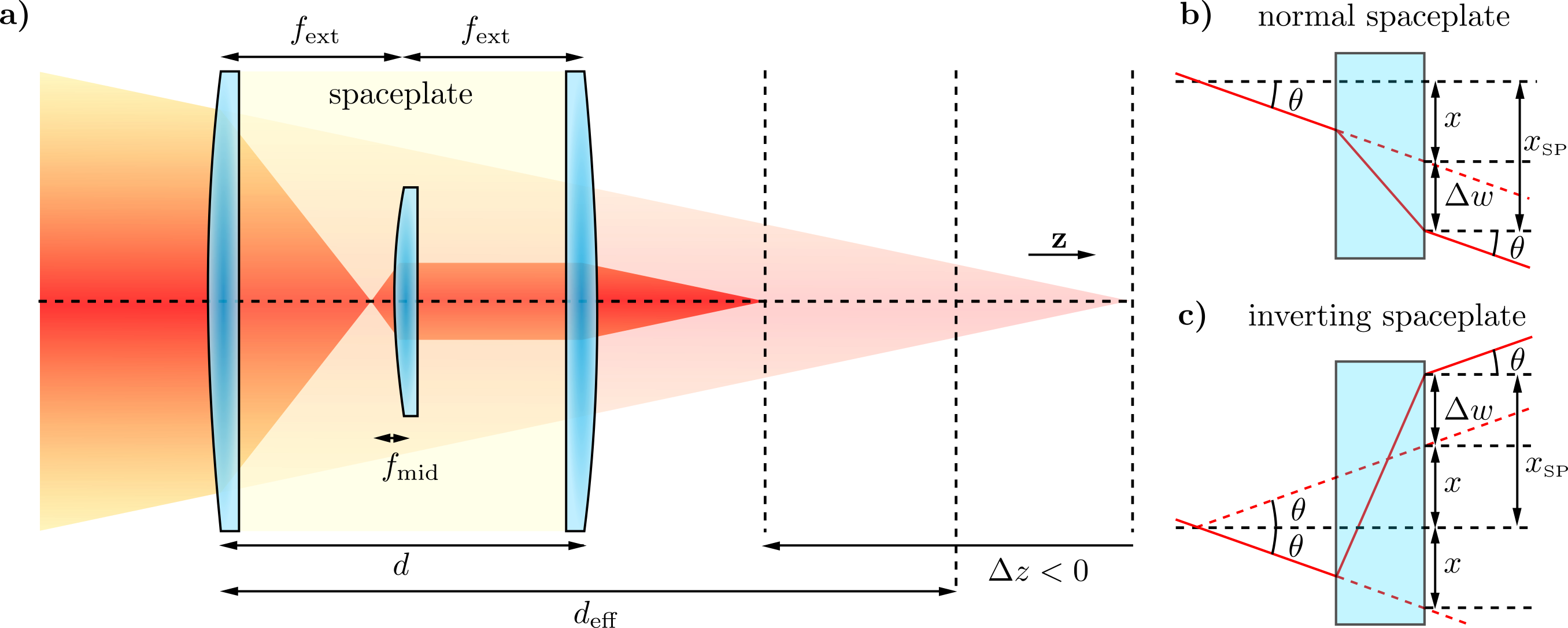}
     \caption{Propagation of a light beam in a three-lens spaceplate. a) Light travels through a spaceplate of thickness $d$ with an equivalent propagation through free-space a distance $d_{\text{eff}}> d$. b) Rays passing through a spaceplate will maintain their angle $\theta$, but are displaced by walkoff $\Delta w$, just as if they were passing through free space. c) Inverting spaceplates do the same while also inverting the rays about the $\mathbf{z}$-axis. The transverse direction is $\mathbf{x}$, and the longitudinal direction is $\mathbf{z}$. In this figure, $f_{\text{ext}}$ is the focal length of the exterior lenses, $f_{\text{mid}}$ is the focal length of the middle lens, $x$ and $x_{\text{SP}}$ are the transverse displacements of the original and space-compressed rays entering at angle $\theta$, respectively, and $\Delta z$ is the longitudinal shift of an output light field.}
     \label{fig:20220922-threeLensDiagram}
\end{figure*}
The three-lens spaceplate (Fig. \ref{fig:20220922-threeLensDiagram}) utilizes the Fourier plane of a 4-$f$ lens system to impart the same angle-dependent phase that free-space propagation imparts. Our 4-$f$ lens system consists of two positive lenses of focal length $f_{\text{ext}}$, the "exterior lenses". The first lens maps the complex amplitude distribution of an incident field across the front focal plane to the back focal plane via a Fourier transform \cite{Goodman2017}. Thus, there is a one-to-one mapping between the angle of an incoming plane wave and the position in the back focal plane, known as the "Fourier plane", which is located a distance $f_{\text{ext}}$ after the lens. The second lens is a distance $2f_{\text{ext}}$ after the first lens and performs another Fourier transform, undoing the action of the first lens. With a spatially-varying phase element (i.e., a "phase mask") in the Fourier plane, the 4-$f$ system phase shifts each incoming plane wave but otherwise leaves it unchanged. The angle-dependent phase shift required to emulate free-space propagation happens to correspond to the phase mask created by a thin positive lens. Consequently, such a lens placed in the Fourier plane (i.e., a middle lens with focal length $f_{\text{mid}}$) will cause the field transmitted by the 4-$f$ system to appear as though it had propagated through a slab of free-space, potentially longer than the 4-$f$ system itself. This three-lens device is similar to a Fourier filtration system which uses a spatial light modulator or a transparency located at the Fourier plane to modify the phase or amplitude of the Fourier decomposition. Note that while it is neither monolithic nor plate-like, we still refer to it as a spaceplate due to its similar effect on light. Note as well that the use of a 4-$f$ system results in inversion of the light about the $\mathbf{z}$-axis ($x=0$); while this does not affect the compression or absolute magnification of the device, it does differentiate the device from other spaceplates.  

In the next section, we theoretically demonstrate how this three-lens design can be used to attain free-space compression on a much larger scale than previous spaceplate designs. In Section \ref{sec:exp}, we experimentally demonstrate the three-lens spaceplate and characterize its performance for a number of lens choices. We show how it can be straightforwardly inserted into the design of long imaging systems, even if they are broadband. In Section \ref{sec:fundamentalLimits}, we outline some fundamental theoretical limits on the performance of the three-lens spaceplate. We focus on compression ratio, NA, and bandwidth, which recent work has identified trade-offs between \cite{Chen2021, Shastri2022, Mrnka2022} for other spaceplate designs. In the last two sections, we compare and contrast the three-lens design to related optical systems, such as the telephoto lens and Cassegrain reflector, and discuss prospects for its application.
\begin{figure*}[!b]
     \centering
     \includegraphics[width=0.8\textwidth]{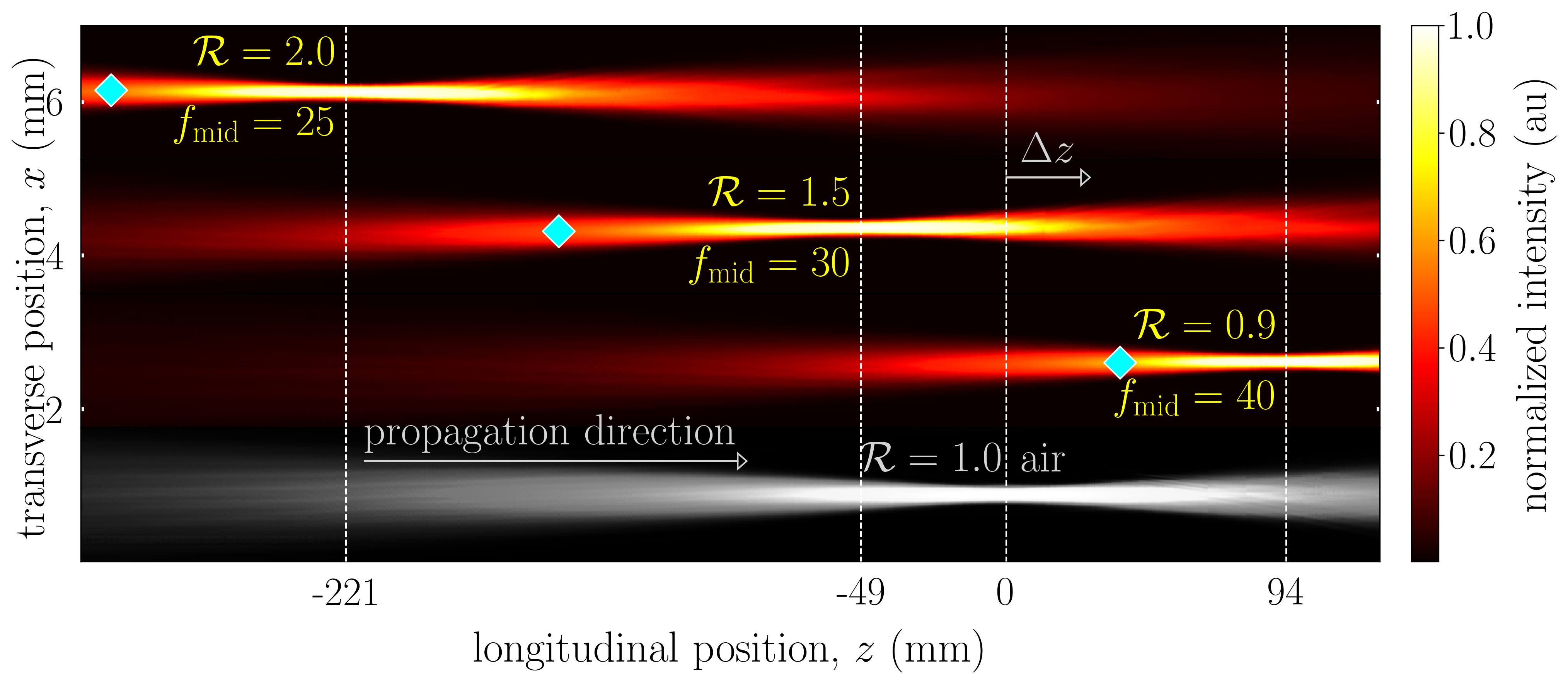}
     \caption{Measurement of the focus advance due to the three-lens spaceplate. At each $z$, we plot the intensity along $x$ (after selecting the $x$ row with the highest intensity) without the spaceplate - just air - (in greyscale, bottom) and for each spaceplate middle lens $f_{\text{mid}}$ (in colorscale, associated with nominal compression ratio $\mathcal{R}$). The vertical dotted lines mark the peak intensity along $z$, i.e. the position of the focus, and the diamond scatter points denote the nominal focus positions. With the air-only focus taken as $z=0$, the vertical lines indicate the focal advance for each spaceplate. Negative advances correspond to a spaceplate that replaces more space than it occupies, i.e., $\mathcal{R} > 1$. For more experimental details and descriptions see Sections \ref{sec:expDesign} and \ref{sec:focusAdvancement}. }
     \label{fig:2022914_focusComparison}
\end{figure*}
\section{Fourier optics description of a three-lens spaceplate}
\label{sec:FourierOptics}
To motivate the design of the three-lens spaceplate, we follow the Fourier optics description of the spaceplate effect introduced by Reshef et al. \cite{Reshef2021}. Suppose we have a light field propagating through free-space centered on the $\mathbf{z}$-axis. We then consider how each transverse spatial Fourier component of this initial field is transformed by this propagation. Each of these components, or plane waves, is described by its momentum vector, $\mathbf{k} = \left(k_{x},k_{z} \right) = k(\sin \theta, \cos \theta)$, where $\theta$ is the angle between the $\mathbf{z}$-axis and the direction of the plane wave. Here, we consider a single transverse dimension ($x$) to reduce notational complexity. This simplification is valid as long as the action of a lens is separable in the $x$ and $y$ directions, which, in the paraxial approximation that we use, is indeed the case. When propagated over a distance $z_0$ in free-space, the amplitude and direction of each plane wave is conserved, but the phase is not. Between two points on the $\mathbf{z}$-axis separated by $z_0$, each plane wave accumulates phase $\varphi = k_{z} z_0 = |\mathbf{k}|\cos(\theta)z_0$. 

A spaceplate must produce this same phase response, but in a shorter distance $d$ than $z_0$. An ideal spaceplate, if inserted into the light path, will produce the phase response $\varphi_{\text{SP}} = k_{z}z_0 = z_0(k^2-k_x^2)^{1/2}$. In the small angle approximation, $\varphi_{\text{SP}} = z_0 k(1-k_x^2/2k^2)$. Then, by neglecting global phase terms, we arrive at our target phase, $\varphi_{\text{SP}} = -z_0 k_{x}^2/2k$. An approach for considering larger incident angles is given in Appendix \ref{sec:appendixLargeAngle}. In the Fourier plane of the 4-$f$ system in Fig. \ref{fig:20220922-threeLensDiagram}(a), the field at position $r$ (i.e., axial distance) is the incoming field component with transverse momentum $k_{x} = rk/f_{\text{ext}}$. Substitution of this relation into the target phase gives the spatial phase mask required to emulate propagation through space, \vspace{-8pt}

\begin{align}
\varphi_{\text{SP}} = -\frac{z_0{kr^2}}{2f_{\text{ext}}^2}.
\label{eq:spaceplatePhase}
\end{align}
The phase added by a spherical, thin lens of focal length $f_{\text{mid}}$ in the paraxial approximation is given by $\varphi_{\text{mid}} = -kr^2/2f_{\text{mid}} $, which is of the same form as Eq. (\ref{eq:spaceplatePhase}). By setting $\varphi_{\text{SP}}=\varphi_{\text{mid}} $ we find that the spaceplate effectively propagates light a distance $z_0 = f_{\text{ext}}^2/f_{\text{mid}}$ . In this way, we can compress space by placing a positive lens at the Fourier plane. 

A slight complication is that, as part of the 4-$f$ system, there already is free-space propagation of distance $f_{\text{ext}}$ before the first lens and $f_{\text{ext}}$ after the last lens. We do not include those regions in the length of space replaced and thus, $d_{\text{eff}}=z_0 -2f_{\text{ext}}$. Similarly, we define the spaceplate thickness to solely be the distance \textit{between} the external lenses, $d=2f_{\text{ext}}$. Therefore, the compression ratio is \vspace{-8pt}
\begin{align}
    \mathcal{R} =\frac{d_{\text{eff}}}{d}= \frac{f_{\text{ext}}}{2f_{\text{mid}}} - 1,
    \label{eq:compressiveRatio1}
\end{align}
which only depends on the focal lengths used. If $\mathcal{R} > 1$ or, equivalently, if $f_{\text{mid}}<f_{\text{ext}}/4$, more space is replaced than is occupied by the three-lens spaceplate. Further discussion of system length, compression ratio, and Eq. (\ref{eq:compressiveRatio1}) is given in Appendix \ref{sec:appendixCompressionRatio}.
\begin{figure}[t]
     \centering
     \includegraphics[width=0.95\textwidth]{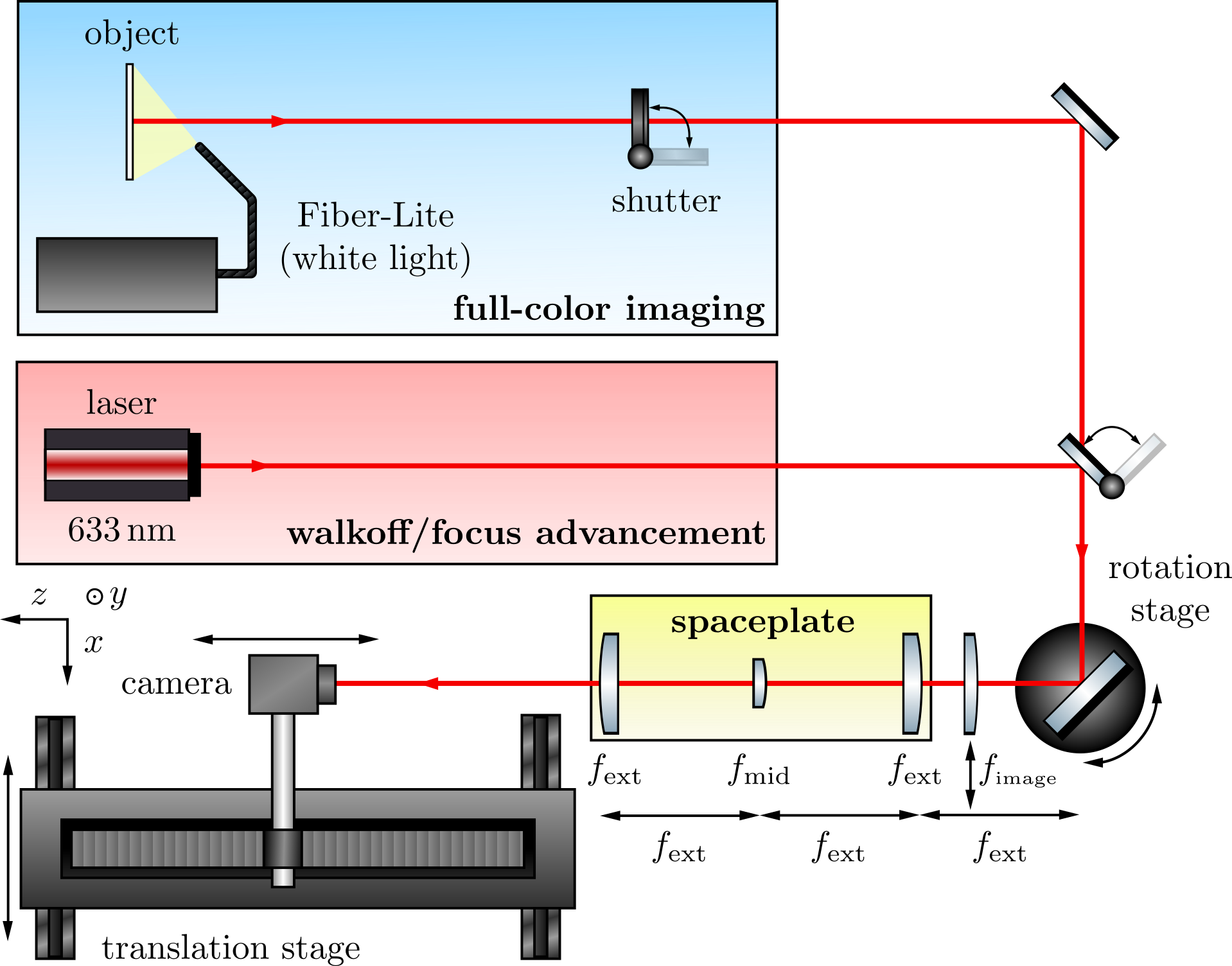}
     \caption{Experimental setup used to test the spaceplate focus advancement, the walkoff, as well as the full-color imaging. The lens labeled $f_{\text{image}}$ is removed for walkoff experiments.}
     \label{fig:expSetUp}
\end{figure}
\section{Experimental demonstration of three-lens space compression}
\label{sec:exp}
\subsection{Design of the focus advancement, walkoff, and imaging experiments}
\label{sec:expDesign}
The setup used to test the three-lens spaceplate is shown in Fig. \ref{fig:expSetUp}. We measure the advance of the focus of a beam, the walk-off of a beam incident at an angle, as well as imaging properties. These three types of measurements require minor setup variations that will be detailed in the following three subsections. 

\begin{figure*}[!b]
     \centering
     \includegraphics[width=0.86\textwidth]{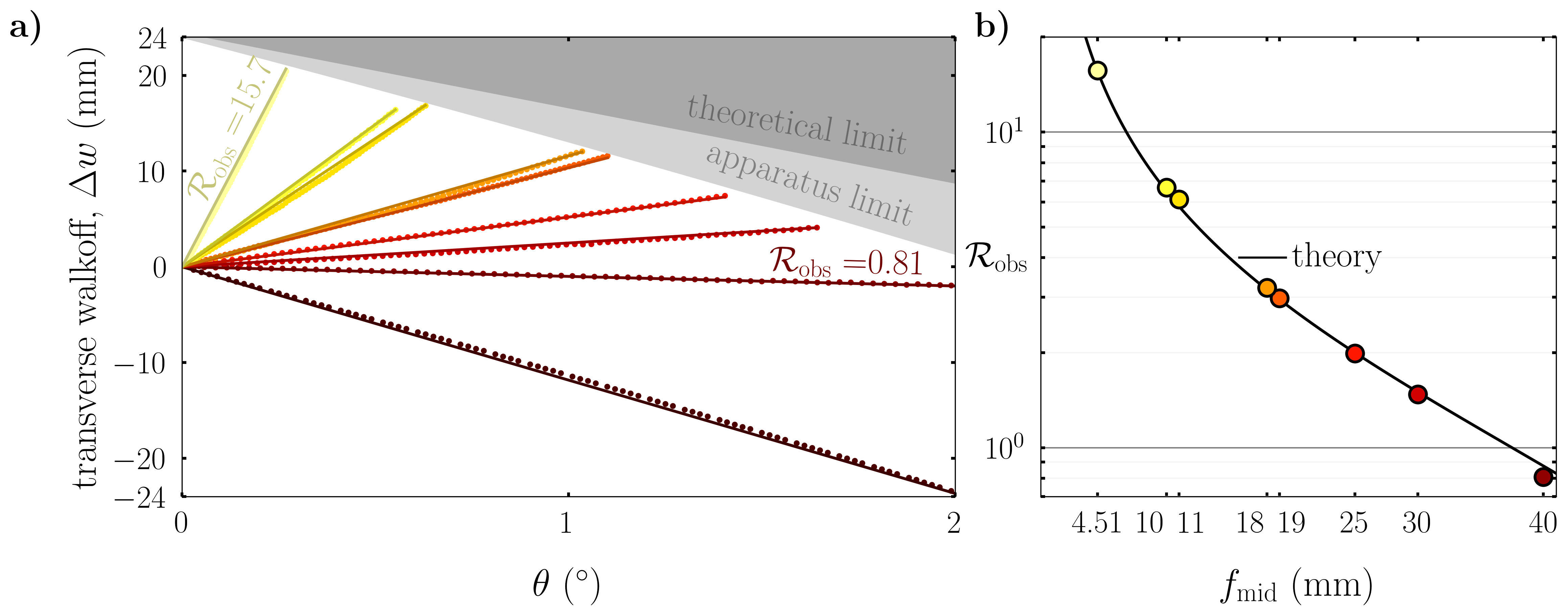}
     \caption{Measurement of the transverse walkoff and compression ratio as a function of middle focal length. a) The measured transverse walkoffs ($\Delta w$) of each of the tested spaceplates are plotted as a function of angle with fitted linear regressions. The theoretical limit is derived from the aperture sizes of the lenses used and the apparatus limit is produced by the bounds of the translation stage and image sensor. b) Here, the compression ratios measured from the corresponding walkoffs in (a) are plotted against the focal lengths of the middle lens ($f_{\text{mid}}$). The point associated with least compression, $f_{\text{mid}}=\infty$, does not appear in b). All points have their uncertainties plotted, though most are obscured by the symbol for the point.}
     \label{fig:2022914_compressiveFactorAndWalkoff}
\end{figure*}
We start by describing elements in the setup common to all three measurements. The external lenses, $f_{\text{ext}}=\SI{150}{mm}$, are the same throughout all the tests and have $\SI{50.8}{mm}$ diameters and $\SI{15.0}{mm}$ thicknesses. We test the effect of the following series of middle lenses: $f_{\text{mid}}= \{40, 30, 25, 19, 18, 11, 10, 4.51\}\,\SI{}{mm}$ whose respective diameters and thicknesses were $\{25.4,25.4,12.7,12.7,6.5,8.0, 5.5\}\,\SI{}{mm}$ and $\{12.5, 14.0, 7.0, 6.0, 2.2,5.0, 6.5, 2.94\}\,\SI{}{mm}$. With the exception of the $f_{\text{mid}}=\{18,11,4.51\}\SI{}{mm}$ aspheric lenses, all lenses used throughout these tests were achromatic doublets, designed and anti-reflective (AR) coated for the visible range (400-$\SI{700}{nm}$ light). The $f_{\text{mid}}= \SI{18}{mm}$ aspheric lens was designed for $\SI{780}{nm}$ light, and the $f_{\text{mid}}= \{11,4.51\}\,\SI{}{mm}$ aspheric lenses were designed for $\SI{633}{nm}$ light. All aspheric lenses were AR-coated for 600-$\SI{1100}{nm}$. After the spaceplate, an image sensor ($\SI{5.86}{\micro m} \times \SI{5.86}{\micro m}$ pixel size, $1936\times1216$ pixels, color) was mounted on a motorized translation stage capable of moving both in the transverse ($x$) and longitudinal ($z$) directions. With this sensor, we recorded $x\times y$ spatial intensity distributions at chosen $z$ positions. The electronic settings of the image sensor were constant throughout the various tests.

\subsection{Focus advancement of various three-lens spaceplates}
\label{sec:focusAdvancement}

 We measure the focus advancement of a $\SI{633}{nm}$ wavelength Helium-Neon laser that is collimated so that its beam has an intensity full-width-half-maximum (FWHM) of $\SI{3.5}{mm}$ using three different middle lens focal lengths corresponding to three distinct compression ratios. To create a focusing beam, the laser travels through a $\SI{50.8}{mm}$ diameter positive lens with a thickness of $\SI{11.3}{mm}$ and a focal length of $f_{\text{image}}=\SI{1000}{mm}$ [see Fig. \ref{fig:expSetUp}]. The three-lens spaceplate is placed between this positive lens and the nominal focus $z$-position, then the image sensor records intensity distributions over a range of $\SI{500}{mm}$ with a $\SI{0.25}{mm}$ step size. Each $x$ pixel row of the recorded distribution is summed to determine the row with the maximal intensity. This row becomes the 1D intensity profile at each $z$, which is plotted in Fig. \ref{fig:2022914_focusComparison}. For each middle lens, the peak intensity along $z$ is taken as the new focal position. 
 
 We measure longitudinal focal shifts of $\Delta z = \SI{+94}{mm}$, $\SI{-49}{mm}$, and $\SI{-221}{mm}$ corresponding to observed compression ratios $\mathcal{R_{\text{obs}}}= 0.69$, $1.16$, and $1.74$, respectively. The respective nominal compression ratios are $\mathcal{R} = 0.9, 1.5,$ and $ 2.0$. While these are significantly different from the measured values, they demonstrate that the three-lens spaceplate can successfully replace space. 

 Deviations between $\mathcal{R}$ and $\mathcal{R_{\text{obs}}}$ result because the three-lens spaceplate relies on a lens located directly at the Fourier plane of the 4-$f$ system, and any deviation of the thick middle lens from this position results in a reduction of compression and non-unity magnification. Specifically, we subsequently observe that upon transmitting an expanded collimated beam with a $\text{FWHM}=\SI{3.5}{mm}$ through the three-lens spaceplate, the resultant beam width varies ($1.7-\SI{4.4}{mm}$) depending on small adjustments ($0-\SI{5}{mm}$) to the longitudinal position of the middle and rear lenses in the spaceplates. This demonstrates that deviations between the expected and observed longitudinal shifts are due to misalignment of the system. The effect of these sensitivities are reduced in subsequent walkoff and imaging experiments.
 
 Nonetheless, the observed negative longitudinal shift demonstrates that this design is capable of replacing free-space on a scale of three orders of magnitude greater than previous spaceplate designs. This experiment is a qualitative demonstration of space compression - the following experiments are used to characterize the three-lens spaceplate with greater accuracy. 
\begin{figure*}[t]
     \centering
     \includegraphics[width=0.8\textwidth]{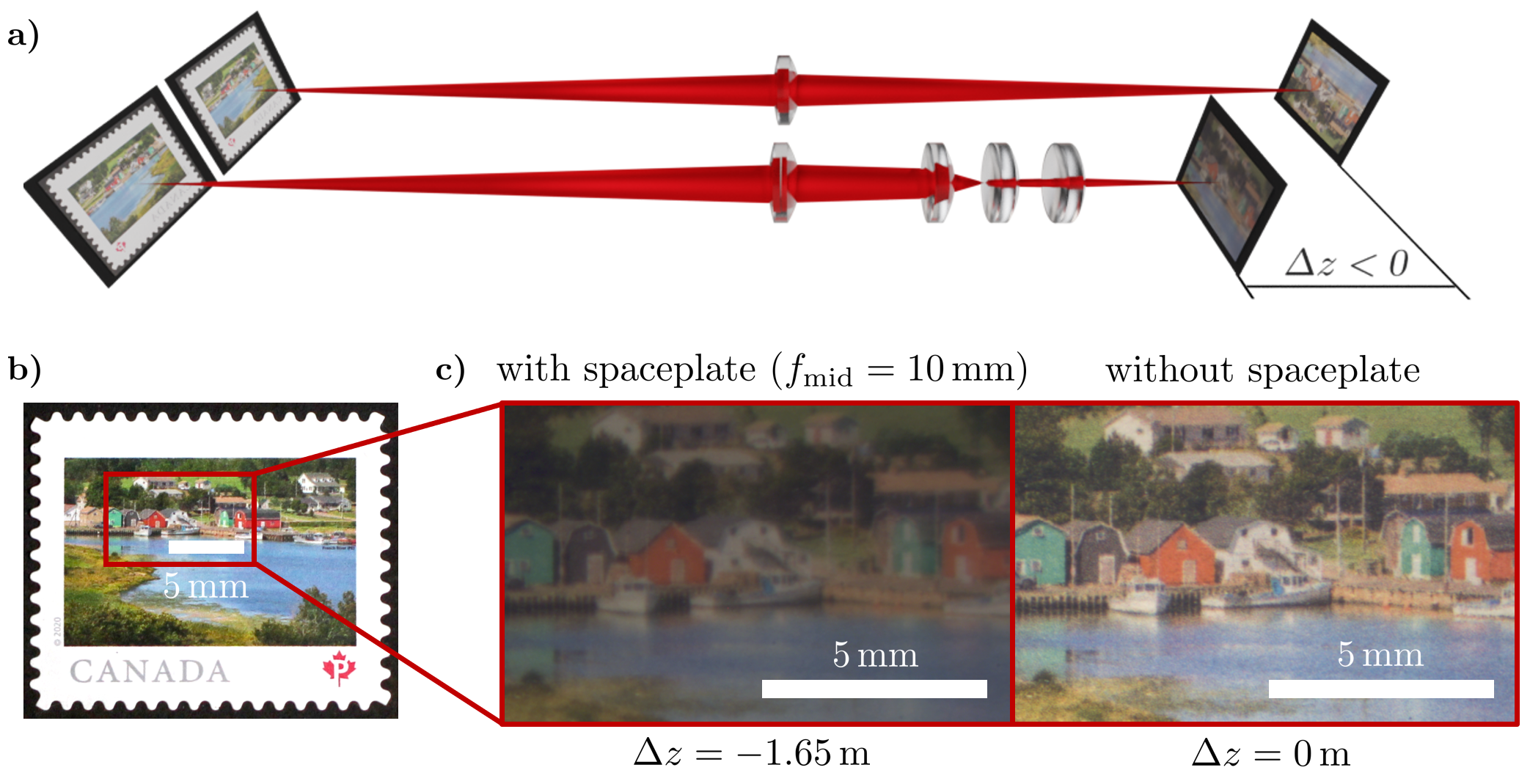}
     \caption{Full-color imaging with a three-lens spaceplate. a) Top: imaging system; Bottom: imaging system with the three-lens spaceplate. b) The imaging object, a stamp; c) An image of the stamp. Right: without the spaceplate; Left: with a $f_{\text{mid}}=10$ mm spaceplate with nominal compression ratio $\mathcal{R}=6.5$ (image is rotated by $180^\degree$ to compensate for the three-lens spaceplate inversion). With the spaceplate, a sharp full-color image is formed $\SI{1.65}{m}$ closer to the imaging lens, compressing the size of the imaging system. The stamp is a photo of the French River in Prince Edward Island by Barrett and Mackay Photography. Reproduction is permitted under the Canada Post Corporation Act \cite{Stamps2009}.}
     \label{fig:20220915_fullColorImageComp}
\end{figure*}

\subsection{Transverse beam walkoff of a three-lens spaceplate}
We next measure the transverse walk-off $\Delta w = |x_{\text{SP}}| - |x| = d(\mathcal{R}-1)\sin{\theta}$ [see Fig. \ref{fig:20220922-threeLensDiagram}(b) for a diagram] of a beam transmitted through the spaceplate, incident at angle $\theta$ to the $z$-axis, rotating about the $y$-axis. For each spaceplate, the walk-off as function of angle characterizes the NA and aberrations, and determines the compression ratio with greater accuracy than in the focal advance measurement. Specifically, before conducting walkoff experiments, the longitudinal positions of the middle and rear lenses are incrementally adjusted until a beam ($\text{FWHM}=\SI{3.5}{mm}$) transmitted through the spaceplate becomes collimated and has unity magnification. That is, until the exiting beam has diameter of $\text{FWHM}=(3.5\pm0.1)\,\SI{}{mm}$ both immediately after the rear lens of the spaceplate, and one meter after it. This process was instituted after observing how sensitive the system was to lens placement during preliminary experiments (Section \ref{sec:focusAdvancement}).

To test the beam walkoff, the imaging lens and beam-expander are absent, leaving the laser beam collimated. The rotation stage in Fig. \ref{fig:expSetUp} varied $\theta$ from 0 to 2 degrees in 0.02 degree increments. The stage is located $f_{\text{ext}}$ before the first lens, and we measure the transverse displacement $z = 4f_{\text{ext}}=\SI{600}{mm}$ from the rotation axis of the stage. The rotation stage is placed here due to our definition of $\text{NA}=\sin\theta$, where $\theta$ is the half-angle of a maximum cone accepted by the system, originating $f_{\text{ext}}$ before the first lens. By putting the stage at this position, we are able to contextualize the experimental angular limitations of the system with an analysis of its numerical aperture. The position at which we measure the transverse displacement, however, does not matter in general. One will measure the same $\Delta w$ regardless of the position of the imaging plane. 

First, we measure the transverse displacement $x(\theta)$ of a beam from the $z$-axis in free space, and then perform analogous measurements $x_{\text{SP}}(\theta)$ of beams transmitted through each spaceplate. Due to inversion, the beams are displaced in opposite directions, but we can still compute the walkoff as: $\Delta w = |x_{\text{SP}}|-|x|$. We measure $x$ and $x_{\text{SP}}$ by fitting Gaussians to the measured transverse intensity profiles. 

Fig. \ref{fig:2022914_compressiveFactorAndWalkoff}(a) presents this series of walkoff measurements for each of the eight middle lenses given in Section \ref{sec:expDesign}, with corresponding observed compression ratios in Fig. \ref{fig:2022914_compressiveFactorAndWalkoff}(b). Each of the configurations produced compression ratios within error of their nominal compression ratios over the entire measured range. For the highest experimental compression ratio, $\mathcal{R}=15.6$, $f_{\text{mid}}=\SI{4.51}{mm}$, the spaceplate replaced 4.39 meters of free-space, which is more than one-thousand times greater than that demonstrated by other spaceplate designs.
\begin{figure*}[t]
     \centering     \includegraphics[width=0.8\textwidth]{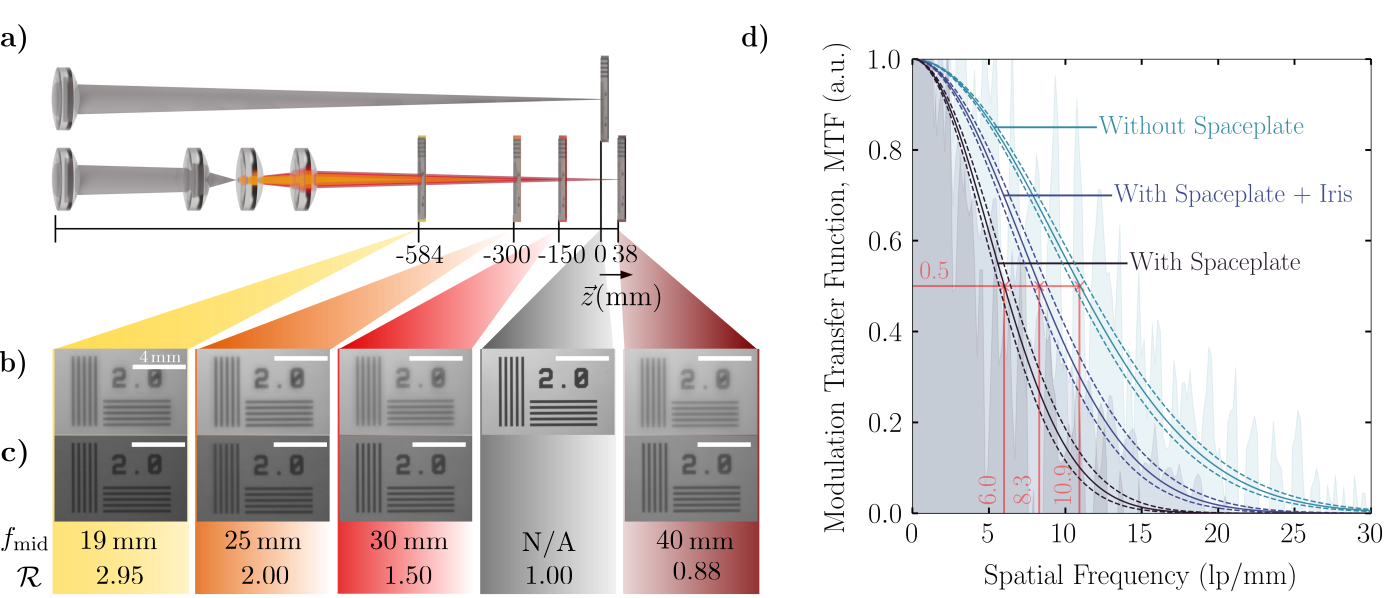}
     \caption{The effect of the three-lens spaceplate on resolution and contrast. a) The three-lens spaceplate is used to compress an imaging system consisting of a $\SI{1000}{mm}$ focal length doublet. b) Images taken using different spaceplates. c) Images taken with a spaceplate and an iris. d) Measured (background, filled plots) and fitted Modulation Transfer Functions (MTF) for three full-color stamp images; one was taken with only the imaging lens, another with the imaging lens and a spaceplate ($\mathcal{R} = 6.5$, $f_{\text{mid}} = \SI{10}{mm}$), and the third with an iris just before the third spaceplate lens. The dotted curves represent the standard error resulting from the method used to approximate the MTFs. See the text for discussion and comparison of results.}
     \label{fig:20221121_monochromaticimagingupdated}
\end{figure*}
\subsection{Broadband imaging with a three-lens spaceplate}
In our final measurement, we investigate the performance of the three-lens spaceplate for imaging with full-visible-spectrum light. We first illuminate a postage stamp by white-light and use the $f_{\text{image}}$ lens to form a sharp image on the image sensor. The object and image distances are approximately $\SI{1.9}{m}$ and $\SI{2.1}{m}$, respectively, and we measure a magnification of 1.1. We record images with and without a $f_{\text{mid}}=\SI{10}{mm}$ spaceplate [Fig. \ref{fig:20220915_fullColorImageComp}(c), left and right panes, respectively] in the imaging path. Before imaging, the middle and rear lenses of the spaceplate are calibrated using an expanded collimated beam, as was done in the walkoff experiment. The camera is then placed at the nominal longitudinal position relative to the focus position seen without a spaceplate in the imaging path. Then, the middle and rear lens longitudinal positions are adjusted slightly to achieve the same magnification seen without the spaceplate. To reduce aberration and edge-effects, a large iris was positioned immediately before the third lens in the spaceplate and was adjusted until the highest contrast and resolution were seen. This was at an aperture size of $(12\pm2)\,\SI{}{mm}$. While there still is a reduction in contrast and brightness, which we discuss later, the images show that the three-lens spaceplate can advance the image plane of a full-visible-spectrum imaging system by 1.65 m. 

Next, we image an NBS 1963a resolution target in order to characterize the reduction in resolution and contrast caused by the addition of four different three-lens spaceplates (see Fig. \ref{fig:20221121_monochromaticimagingupdated}). The addition of the spaceplates does not change the magnification, as expected. However, Fig. \ref{fig:20221121_monochromaticimagingupdated}(b) shows that every spaceplate reduces image contrast and resolution in comparison to the image formed in the absence of a spaceplate. We quantify this reduction with the modulation transfer function (MTF) of the imaging system following the method of Saiga et al. \cite{Saiga2018} in Fig. \ref{fig:20221121_monochromaticimagingupdated}(d). For the case of $f_{\text{mid}} = \SI{10}{mm}$ ($\mathcal{R} = 6.5$) and the stamp imaged without an iris, the spaceplate lowers the spatial frequency at half-max of the MTF by 45\% (i.e. $10.9\,\text{lp/mm}$ is reduced to $6.0\,\text{lp/mm}$). Three potential causes for this reduction are aberrations, scattering from lens edges, and diffraction by the lens apertures. The first two would be mitigated by a reduced aperture, whereas the effect of diffraction would be worsened. Adjusting the diameter of an iris $[(12\pm2)\,\SI{}{mm}]$ just before the third lens, we take the images shown in Fig. \ref{fig:20221121_monochromaticimagingupdated}(c), which are qualitatively improved in contrast and resolution. This improvement is reflected in the MTF measured with an iris in place, which increases by 38\% (i.e. $6.0\,\text{lp/mm}$ to $8.3\,\text{lp/mm}$) suggesting that diffraction is not the main cause of the reduction. Equivalently stated, addition of the spaceplate and iris causes a 24\% reduction in the MTF (i.e. $10.9\,\text{lp/mm}$ to $8.3\,\text{lp/mm}$).

\section{Fundamental limits on performance of the three-lens spaceplate}
\label{sec:fundamentalLimits}

In this section, we briefly discuss limits on the NA and compression inherent to the three-lens design of the spaceplate. In most physically realizable setups, the NA of the three-lens spaceplate will be limited by the diameter of the third lens, $D_3$. For this common case, the NA can be expressed in terms of the f-number of the third lens, $N_3 = f_{\text{ext}}/D_3$, and $\mathcal{R}$:

\begin{align}
    \text{NA} = \left[1+4N_3^2\left(1+2\mathcal{R}\right)^{2}\right]^{-\frac{1}{2}}.
    \label{eq:numericalaperturethirdlens}
\end{align}
Eq. (\ref{eq:numericalaperturethirdlens}) shows that there is trade-off between compression ratio and NA. The lower the f-number of the third lens, the less strict this trade-off is. The third lens used in our experiment has $N_3=3$, which limits the NA of the three-lens spaceplate be less than 0.06, worsening as $\mathcal{R}$ increases. This represents the greatest shortcoming of the three-lens spaceplate design: even when using a relatively large third lens diameter ($D_3=\SI{50}{mm}$), the maximum incident angle is three degrees from the optical axis. For information on the less common cases where the first or second lenses limit the NA, see Appendix \ref{sec:appendixDiffraction}.

We have considered two additional physical mechanisms inherent to the three-lens spaceplate that lead to limitations. However, these turn out to be much less restrictive than the above limitation in practical applications with standard lenses. Firstly, the Abbe Sine Condition limits the NA independent of the lens diameters. Secondly, diffraction from the first lens in the spaceplate limits the achievable compression ratio. These two limitations may be relevant for new types of lenses, such as metalenses, and are therefore explored in depth in Appendices \ref{sec:appendixAbbe} and \ref{sec:appendixDiffraction}, respectively. 

\section{Comparison to other compressive optics}
A number of distinctions must be made between the three-lens spaceplate and other methods to conserve space in imaging systems. We start by comparing it to other spaceplate designs. All of these exhibit some degree of transverse translation invariance \cite{Shastri2022}, whereas the three-lens spaceplate does not, since lenses are not translationally invariant. Another difference is that the three-lens spaceplate  inverts the transmitted field through the axis of the lenses (i.e., $x\rightarrow-x$ and $y\rightarrow-y$), though this is avoided simply by using a pair of spaceplates.

Many conventional imaging systems are already designed to minimize the space they occupy. Examples include the telephoto lens and the Cassegrain reflector whose goal is to create an imaging system shorter than its effective focal length $f_{\text{eff}}$. Thus the compression of space is intertwined with the main purpose of the imaging system - to provide optical magnification. In practice, telephoto lenses are limited to a length $> 0.8 f_{\text{eff}}$ \cite{Kingslake1978, Tremblay2005, Tremblay2007}. This is equivalent to a non-telephoto lens followed by a three-lens spaceplate with $\mathcal{R}\geq 1/0.8=1.25$, much lower than the highest value reported here for three-lens spaceplates $\mathcal{R}=15.7$, albeit with a higher NA.

The nature of the components of these three-lens systems means that they are large. In this way, they differentiate themselves from thin spaceplates. Because they use lenses, implementable three-lens spaceplates have limitations on thickness, aperture, and, consequentially, compression ratio. Using off-the-shelf components and manageable system lengths ($2f_{\text{ext}}\lesssim\SI{1}{m}$), compression ratios are limited ($\mathcal{R}\lesssim 20$); however, due to the large size of these systems, they can save large absolute quantities of space (on the order of meters). This is in contrast to existing spaceplate designs, which only save space on the order of microns or millimeters \cite{Reshef2021, Guo2020, Page2022, Chen2021}. 

One advantage of using standard optical components in three-lens spaceplates is that the resulting spaceplate is broadband, polarization-independent, and highly configurable in compression ratio. The compression ratio of the three-lens spaceplate can easily be changed by replacing the middle lens, whereas the compression ratio of other spaceplate designs is fixed upon manufacturing.

Similar systems have been used in microscopy to rapidly tune imaging planes \cite{Zuo2013, Chen2018, Kang2020, Zuo2015} in transport-of-intensity phase microscopy or to add free space to optical systems \cite{Schulze2013}. However, such implementations do not compress space and are instead intended to artificially increase the depth of field. 

\section{Conclusion}

In this paper, we have successfully demonstrated the compression of free-space using a device that we call a "three-lens spaceplate", a positive lens placed at the Fourier plane of a 4-$f$ optical setup. We have shown that our spaceplate design can be used for meter-scale space compression and the miniaturization of broadband imaging systems. Using the three-lens spaceplate, we have achieved experimental compression ratios up to $\mathcal{R}_{\text{obs}}=15.7$ that replace 4.4 meters of free-space. Until now, replication of the transfer function of free-space over lengths of this scale had not been demonstrated: previously, space compression of only one or two millimeters had been shown \cite{Reshef2021}. The three-lens spaceplate has not only shown that this is possible, it has accomplished this using off-the-shelf optical components. Moreover, the system was shown to reduce the length of both monochromatic and full-color imaging systems with a 24\% loss of resolution as quantified using its modulation transfer function. Finally, we discussed the trade-off between compression ratio and numerical aperture. Specifically, we found that the numerical aperture is highly limited for the three-lens spaceplate and represents its greatest shortcoming. 

The simplicity and versatility of this spaceplate design motivate its application and future development, though there are certain limitations which differentiate its applications from other spaceplates. The three-lens spaceplate represents a simple, cost-efficient method for reducing the size of axially-symmetric, long focal length systems such as telescopes. Because these three-lens spaceplates are so large and have such limited numerical apertures, however, their application in smaller devices such as smartphone cameras or virtual reality headsets would be difficult to implement. That said, the inherent use of free-space within the system itself implies the potential for even higher compression ratios through the use of metalenses and other space-saving devices. Metalenses could potentially also reduce the numerical aperture limitations of the three-lens spaceplate by reducing lens thickness, and increasing aperture size. Future work includes using such hybrid designs to reduce overall system length and increase numerical aperture, as this could make the implementation of three-lens spaceplates in small devices more realistic.

\subsection*{Funding}
The authors acknowledge support from the Canada Research Chairs Program, the Natural Sciences and Engineering Research Council of Canada, as well as the Canada First Research Excellence Fund. 
\subsection*{Acknowledgments}
The authors would like to thank Orad Reshef, Yaryna Mamchur, and Ryan Hogan for encouragement and discussions, as well as Cheng Li and Aldo Martinez for experimental lendings and accommodations. 

\subsection*{Disclosures}
The authors declare no conflicts of interest.

\subsection*{Data availability}
Data underlying the results presented in this paper are not publicly available at this time but may be obtained from the authors upon reasonable request.

\printbibliography

\newpage

\appendix

\section{Performance beyond the small-angle regime}
\label{sec:appendixLargeAngle}

In motivating the design of the three-lens spaceplate (see Section \ref{sec:FourierOptics}), we used the small-angle approximation to draw similarities between the ideal spaceplate phase, $\varphi_{\text{SP}}$, and the phase of a spherical, thin lens $\varphi_{\text{mid}}$. In reality, this approximation is more restrictive than necessary for such a comparison and thus the possibility of extending beyond this regime merits further discussion.

The general spaceplate phase is $\varphi_{\text{SP}} = kd_{\text{eff}}(1-k_x^2/k^2)^{1/2}$, while the spherical lens phase is $\varphi_{\text{mid}} = k(f_{\text{mid}}^2-x^2)^{1/2}$. Assuming the first lens is capable of performing a perfect Fourier transform of the incident field such that $k_{x} = rk/f_{\text{ext}}$ holds for large angles, we can directly compare the spaceplate phase to that of a spherical lens outside the small-angle approximation. The maximum value of $k_{x}/k=\sin{\theta}$ such that the phase difference is less than some fraction of a period (i.e. $\sim\pi/5$), disregarding any global phase, determines the maximum NA of the three-lens spaceplate. Geometrically, it represents the angle at which the circular phase profile of a spherical lens deviates a fraction of a cycle from the elliptical phase profile of an ideal spaceplate. This method can be generalized to other middle lens forms as long as the first lens linearly maps each incident transverse wavevector component $k_{x}$ to a single position $x$ in the back focal plane.

It must be said, however, that conventional lenses produce limits on NA that make this method of no practical use. As we have shown for our setup, this is because the limits on NA will always be more restrictive than those predicted by the method above. There are likely lens designs that are more suited for matching the phase profile of an ideal spaceplate in which case this method may prove useful.

\section{Defining the compression ratio and system length}
\subsection{Ray-matrix approach to deriving the compression ratio}
\label{sec:appendixCompressionRatio}

One can also assess the three-lens spaceplate using a ray matrix method \autocite{Goodman2017,Macukow1983, Arsenault1983,Xiyuan2008}. The ray-transfer matrix for the system is given by \vspace{-8pt}

\begin{align*}
    M = -\begin{bmatrix}
    \hspace{3pt}1 & \frac{f_{\text{ext}}^2}{f_{\text{mid}}} - 2f_{\text{ext}}\hspace{4pt}\\
    \hspace{3pt}0 & 1\hspace{3pt} 
    \end{bmatrix}
\end{align*}
which is equivalent to the ray-matrix for inversion and propagation over an effective distance $d_{\text{eff}}={f_{\text{ext}}^2}/{f_{\text{mid}}} - 2f_{\text{ext}}$. The compression ratio can then be expressed as\vspace{-8pt}

\begin{align*}
    \mathcal{R} = \left(\frac{f_{\text{ext}}^2}{f_{\text{mid}}} - 2f_{\text{ext}}\right)\frac{1}{d} = \frac{f_{\text{ext}}}{2f_{\text{mid}}} - 1
\end{align*}
which is identical to Eq. (\ref{eq:compressiveRatio1}).

\subsection{The effect of the defined system length}

In the initial Fourier optics derivation of the spaceplate compression ratio (Section \ref{sec:FourierOptics}), we consider a 4-$f$ system without a thickness $f_{\text{ext}}$ of free-space before or after the external lenses $(d=2f_{\text{ext}})$. It is mathematically simpler to consider a system with these preceding and succeeding lengths of free space $(d=4f_{\text{ext}})$; this is because they are required to perform a Fourier transform without any additional phase factors \autocite{Goodman2017}. In such a system, we can directly equate the spaceplate phase [Eq. (\ref{eq:spaceplatePhase})] with the lens phase $(\varphi_{\text{SP}}=\varphi_{\text{mid}})$ and solve for the compression ratio $\mathcal{R}_{4f_{\text{ext}}}$, where we use a subscript to indicate the redefinition of system length: \vspace{-8pt}

\begin{align}
    \mathcal{R}_{4f_{\text{ext}}} = \frac{f_{\text{ext}}}{4f_{\text{mid}}}.
    \label{eq:compressiveRatio2}
\end{align}
This is similar to the initial definition of compression ratio [Eq. (\ref{eq:compressiveRatio1})] but it predicts less compression due to the added free space.

In comparing the two definitions of compression ratios, one may notice that Eq. (\ref{eq:compressiveRatio1}) can evaluate to be negative, but this is simply a demonstration that the three-lens spaceplate can extend the length of systems, not just compress them. Recall that the compression ratio is defined as $\mathcal{R} = d_{\text{eff}}/d$ and that negative compression ratios simply define a system for which $d_{\text{eff}}<0$. Effectively, such systems produce an output field equivalent to a field seen a distance $d_{\text{eff}}$ before the first lens. This is practically irrelevant to our exploration as only compression ratios greater than unity are of interest, though an example can be used to illustrate this discussion as well as demonstrate that Eqs. (\ref{eq:compressiveRatio1}) and (\ref{eq:compressiveRatio2}) are physically equivalent.

Suppose we have a three-lens spaceplate without a middle lens ($f_{\text{mid}}\rightarrow \infty$) such that $\mathcal{R} = -1$ and $\mathcal{R}_{4f_{\text{ext}}} = 0$, as given by Eqs. (\ref{eq:compressiveRatio1}) and (\ref{eq:compressiveRatio2}). Eq. (\ref{eq:compressiveRatio1}) predicts that the light field immediately after the final lens will equal - ignoring inversion - the field a distance $2f_{\text{ext}}$ before the first lens in the spaceplate. This is a separation of $4f_{\text{ext}}$ between equivalent fields. Further, Eq. (\ref{eq:compressiveRatio2}) predicts that the field at the output (a distance $f_{\text{ext}}$ past the final lens in the spaceplate) will equal the field a distance $f_{\text{ext}}$ before the first lens. This is also a separation of $4f_{\text{ext}}$ between equivalent fields. This demonstrates that both compression ratio definitions [Eqs. (\ref{eq:compressiveRatio1}) and (\ref{eq:compressiveRatio2})] are contextually equivalent.

\section{Fundamental Limits}
\subsection{NA limitations due to lens diameter}
\label{sec:appendixNA}
\begin{figure}[h]
    \centering
    \includegraphics[width = \textwidth]{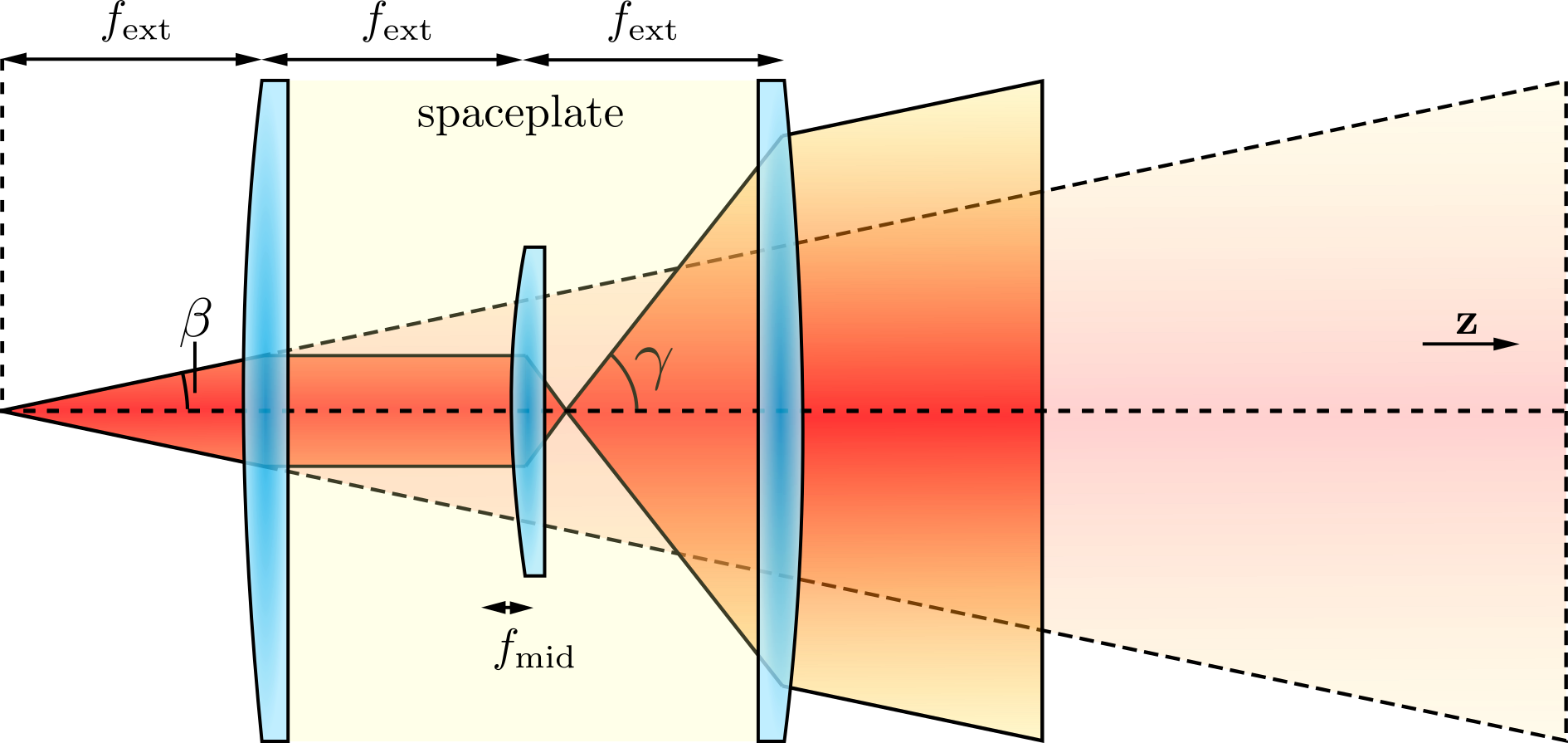}
    \caption{Maximal light cone used to define the numerical aperture (NA) of the three-lens spaceplate and contextualize the Abbe Sine Condition (ASC). We define $\text{NA}=\sin\beta$ for a boundary ray at angle $\beta$ from the $z$-axis, crossing the $z$-axis a distance $f_{\text{ext}}$ before the first lens. See text for discussion of NA and the ASC limitations.}
    \label{fig:threeLensDiagramWalkoff}
\end{figure}
\begin{figure*}[!b ]
     \centering
     \includegraphics[width=0.72\textwidth]{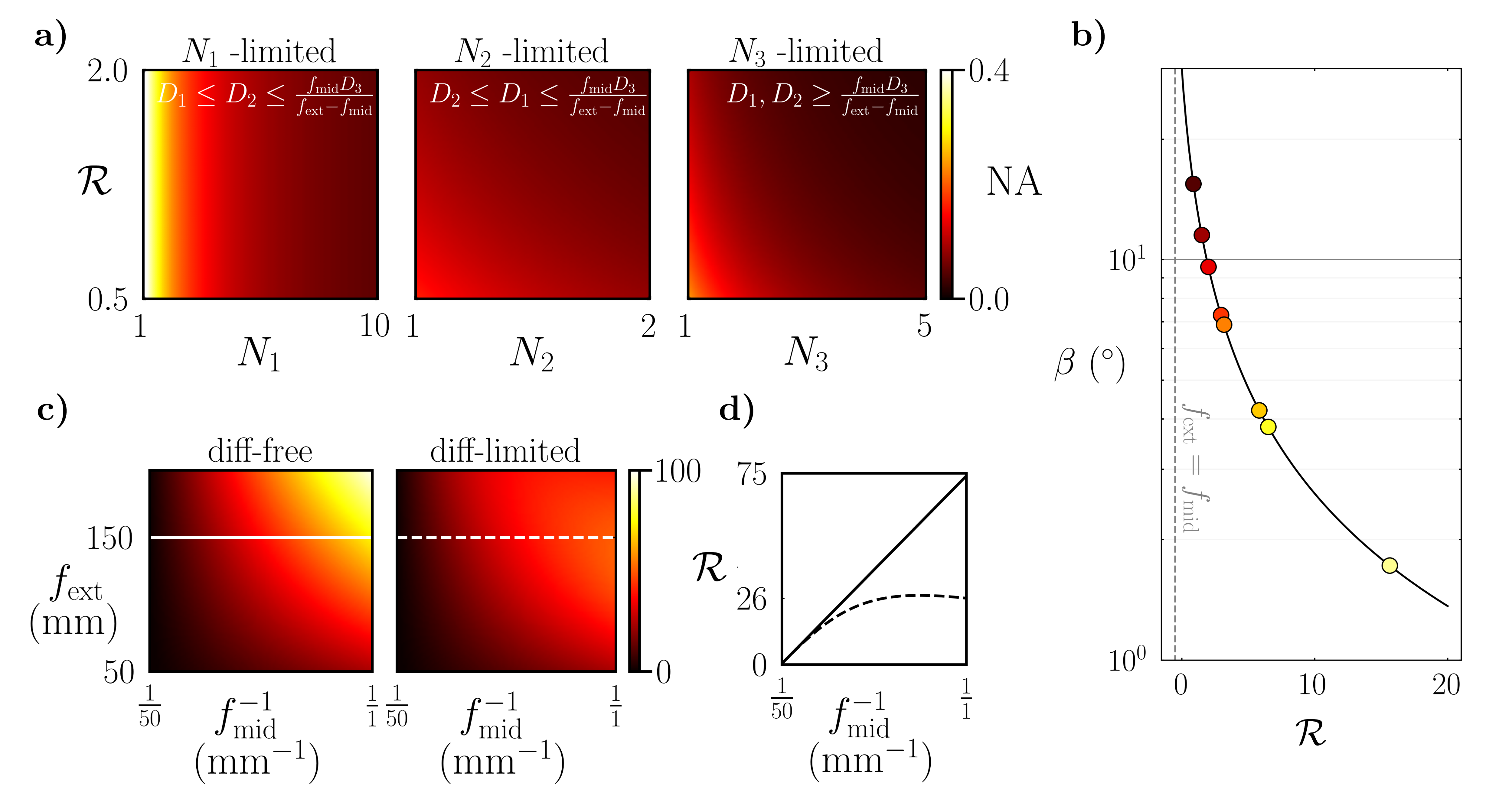}
     \caption{Various fundamental limitations of three-lens spaceplates. a) These three plots show the three cases in Eq. (\ref{eq:numericalaperture}) labeled by the aperture which limits the system NA. b) A plot of the relationship between $\mathcal{R}$ and the maximum input angle as given by the Abbe Sine Condition: each scatter point gives the limit for each experimentally tested spaceplate. c) This plot compares the compression ratio as a function of focal lengths $f_{\text{ext}}$ and $f_{\text{mid}}$ as predicted by Eq. (\ref{eq:compressiveRatio2}) to that as predicted by Eq. (\ref{eq:diffractionLimCompressionRatio}) which considers diffractive effects. In order to make the effect more apparent, the diameter of the first aperture is made artificially small ($\sim \SI{10}{mm}$). d) This plot shows a line cut of both plots in (c) for our experimentally tested focal lengths. The solid line denotes the original expression, and the dashed line denotes the diffraction-limited expression.}
     \label{fig:2022916_limitationsPlotCombined}
\end{figure*}
The use of finite-sized lenses results in a number of limitations which must be considered. The first limitation we discuss is that of NA which depends on the system parameters including compression ratio and the f-numbers of the three-lenses used, $N_1 = f_{\text{ext}}/D_1$, $N_2=f_{\text{mid}}/D_1$, and $N_3=f_{\text{ext}}/D_3$, where $D_i$ is the diameter of the $i$-th lens. As before, we define $\text{NA}=\sin\beta$ for a boundary ray at angle $\beta$ from the $z$-axis, crossing the $z$-axis a distance $f_{\text{ext}}$ before the first lens [see Fig. (\ref{fig:threeLensDiagramWalkoff})]. The angular range of the spaceplate is defined as the largest angle $\beta$ at which a ray can successfully be transmitted through the spaceplate. In most cases, the third lens limits the angular range of the spaceplate as evident in Fig. (\ref{fig:threeLensDiagramWalkoff}), where we assume $f_{\text{ext}}>f_{\text{mid}}$ such that the resulting compression ratio is greater than one. This is because as long as the system compresses space, the third lens must act on a range of angles greater than that incident on the system. However, if the leading or middle lens diameter is reduced enough, the overall limitation on NA will no longer be set by the third lens and instead will be set by the first or second lens. Considering these different cases, we arrive at the following expressions for the NA:\vspace{-8pt}

\begin{align}
    \text{NA}=\begin{cases}
\left(1+4N_{1}^{2}\right)^{-\frac{1}{2}}, & D_{1}\leq D_{2}\leq \frac{f_{\text{mid}}D_{3}}{f_{\text{ext}}-f_{\text{mid}}}\\
\left[1+16N_{2}^{2}\left(1+\mathcal{R}\right)^{2}\right]^{-\frac{1}{2}}, & D_{2}\leq D_{1}\leq \frac{f_{\text{mid}}D_{3}}{f_{\text{ext}}-f_{\text{mid}}}\\
\left[1+4N_3^2\left(1+2\mathcal{R}\right)^{2}\right]^{-\frac{1}{2}}, & D_{1},D_{2}\geq \frac{f_{\text{mid}}D_{3}}{f_{\text{ext}}-f_{\text{mid}}}.
\end{cases}
    \label{eq:numericalaperture}
\end{align}
Eq. (\ref{eq:numericalaperture}) is plotted in Fig. (\ref{fig:2022916_limitationsPlotCombined}a) where each piece-wise function is plotted independently. The third case, where the system is $N_3$-limited, is most physically realizable. The right-most figure in Fig. (\ref{fig:2022916_limitationsPlotCombined}a) shows that for compression ratios greater than one, and $N_3=3$ as used in our experiment, the achievable NA is less than 0.06.

\subsection{NA limitations due to the Abbe Sine Condition}
\label{sec:appendixAbbe}

In designing high-quality imaging systems using common refractive optics, the minimization of aberrations is of great importance. Aplanatism, or the absence of spherical aberration and coma, may be achieved by satisfying two conditions: the system must be axially stigmatic (fully corrected for on-axis imaging) and the system must satisfy the Abbe Sine Condition (ASC) \autocite{Burge2010}. For a system to produce sharp images of extended flat objects, which was a central objective of this work, the latter is of particular interest and thus merits further discussion. 

As we have shown, the three-lens spaceplate does not provide any magnification or lensing effects and hence is not an imaging system, however, the ASC is still relevant. We can see from Fig. (\ref{fig:threeLensDiagramWalkoff}) that the spaceplate can be divided into two subsystems: a leading imaging system of magnification $\mathcal{M}$ and an auxiliary imaging system of reciprocal magnification ${\mathcal{M}}^{-1}$. For simplicity, we assume an object placed a distance $f_{\text{ext}}$ from the first lens and that each imaging system is free of spherical aberration. In other words, we will only examine limitations on NA due to satisfying the ASC.

We begin by defining the magnification of each imaging system. According to geometrical optics, the transverse magnification constant for the leading imaging system is given by: \vspace{-8pt}

\begin{align}
    \mathcal{M} = -\frac{f_{\text{mid}}}{f_{\text{ext}}} = -\frac{1}{2(\mathcal{R}+1)}.
    \label{eq:magnificationconstant}
\end{align}

As we are only interested in systems with compression ratios greater than unity, $\mathcal{R}>1$, the magnification constant for the leading imaging system will always be less than ${1}/{4}$. In contrast, the magnification of the auxiliary imaging system will always be greater than four. Therefore, the leading imaging system is responsible for creating the largest internal ray angles [set by $\gamma$ in Fig. (\ref{fig:threeLensDiagramWalkoff})] and, hence will limit the incident angles according to the ASC: $\mathcal{|M|} = {\sin{\beta}}/{\sin{\gamma}}$ \autocite{Hecht2017}. This condition states that the ratio of the sine of the angle made by a ray exiting the object of an imaging system to that of the ray at the imaging plane must be equal to the system's magnification constant - where either angle is measured from the optical axis as in Fig. (\ref{fig:threeLensDiagramWalkoff}). Using Eq. (\ref{eq:magnificationconstant}), the ASC gives the maximum incident angle the three-lens spaceplate can act on irrespective of aperture size limitations: $\sin{\beta}_{\text{max}} = [2(\mathcal{R}+1)]^{-1}$. For practical lenses, this limit is not attainable since the effects of finite lens diameter are more restrictive.

\subsection{Compression ratio limitations due to diffractive effects}
\label{sec:appendixDiffraction}

Another limitation set by finite aperture size is the diffractive limit on the maximum compression ratio achievable by the spaceplate. For simplicity, we consider diffraction from only the first lens the phase imparted by the middle lens. This assumption is valid for most realistic cases in which the first and third lens are approximately the same size such that the additional diffraction effects attributable to the third lens are negligible. The aperture function of the first lens is Fourier transformed by the first lens, and the resultant point spread function is optically convolved with the middle lens phase [Eq. (\ref{eq:spaceplatePhase})] at the Fourier plane. We approximate the point spread function of the first aperture by a Gaussian with standard deviation $\sigma = g \lambda f_{\text{ext}}/2D_1$ where $g=1.22$ is the Airy first node scaling, $D_1$ is the diameter of the first lens, and $\lambda$ is the wavelength: \vspace{-8pt}

\begin{align*}
    H_1 = \frac{1}{\sqrt{2\pi\sigma^2}}\exp\left( \frac{-x^2}{2\sigma^2} \right). 
\end{align*}
The phase distribution across the range of input $k_x$ is then given by the convolution of the point spread function and the phase of the middle lens.
\begin{align}
\begin{split}
    H\ast l &= \frac{1}{2\pi \sigma^2}\frac{1}{\sqrt{\frac{2a_1^2}{f_{\text{ext}}^2g^2\lambda^2 \pi} + \frac{1}{f_{\text{ext}}\lambda}}} \exp \left( \frac{2a_1^2\pi x^2}{j2a_1^2f_{\text{mid}} \lambda - f_{\text{ext}}^2g^2\pi \lambda} \right) \nonumber\\ 
    &= \underbrace{(1+\alpha^2)^{-1/4} \exp\left( -\frac{\alpha}{1 + \alpha^2} \frac{\pi x^2}{f_{\text{mid}} \lambda}   \right)}_{\text{attenuation}} \\& \qquad \times \underbrace{\exp\left( j\frac{\arctan \alpha}{2}  \right)}_{\text{global phase}} \underbrace{\exp\left( -j \frac{1}{1 + \alpha^2}  \frac{\pi x^2}{f_{\text{mid}} \lambda} \right)}_{\text{phase}} \nonumber
    \end{split}
\end{align}
where $\alpha = 2\pi \sigma^2/f_{\text{mid}}\lambda = g^2 \pi f_{\text{ext}}^2 \lambda/2f_{\text{mid}}a_1^2$. Here, we can ignore the effects of the global phase and radial attenuation, and extract the compression ratios from this new phase. Recognizing the spaceplate phase as $\varphi_{\text{SP}} = -\pi x^2/f_{\text{mid}}\lambda (1+\alpha^2)$, we can equate this to Eq. (\ref{eq:spaceplatePhase}) to solve for the compression ratio:
\begin{align}
    \mathcal{R}_{4f_{\text{ext}}} = \frac{1}{1+\alpha^2} \frac{f_{\text{ext}}}{4f_{\text{mid}}} \Rightarrow \mathcal{R} = \frac{2f_{\text{ext}}f_{\text{mid}}}{4f_{\text{mid}}^2 + \alpha^2f_{\text{ext}}^4} - 1. 
    \label{eq:diffractionLimCompressionRatio}
\end{align}
Note that as the aperture of the first lens gets larger, $\alpha$ goes to zero, and the compression limit approaches our original expression [Eq. (\ref{eq:compressiveRatio2})]. Eq. (\ref{eq:diffractionLimCompressionRatio}) is plotted in Fig. (\ref{fig:2022916_limitationsPlotCombined}c) and (\ref{fig:2022916_limitationsPlotCombined}d): in c), 2D color plots compare the compression ratios achieved given a pair of focal lengths $f_{\text{ext}}$ and $f_{\text{mid}}$, and it is clear that diffraction results in a maximally achievable compression ratio. For example, in d) we set $f_{\text{ext}} = \SI{150}{mm}$ to match our experimental parameters and then sweep $f_{\text{mid}}^{-1}$. As expected, Eq. (\ref{eq:compressiveRatio2}) goes linearly with $f_{\text{mid}}^{-1}$ where as the diffractive considerations result in a global compression ratio maximum. Due to the achievable focal lengths and aperture sizes, however, one would be hard pressed to observe this effect in a real system: Fig. (\ref{fig:2022916_limitationsPlotCombined}c) assumes an initial lens diameter of $\SI{10}{mm}$ which is considerably smaller than achievable lens diameters. Even then, one would need middle lenses with focal lengths smaller than two or three millimeters to observe this limitation in compression ratio. This effect is easily designed around.

\end{document}